\documentclass[a4paper]{article}
\pdfoutput=1

\usepackage{INTERSPEECH2021}

\usepackage{url}
\usepackage[hidelinks]{hyperref}
\usepackage[utf8]{inputenc}
\usepackage{caption}
\usepackage{graphicx}
\graphicspath{ {./figs/} }
\usepackage{amsmath}
\usepackage{booktabs}
\usepackage{multirow}
\title{VRAIN-UPV MLLP's system for the Blizzard Challenge 2021}

\name{Alejandro Pérez-González-de-Martos, Albert Sanchis and Alfons Juan}
\address{
  Machine Learning and Language Processing (MLLP) research group\\
  Valencian Research Institute for Artificial Intelligence (VRAIN)\\
  Universitat Politècnica de València, Spain
}
\email{
  \{alpegon2,josanna2,ajuanci\}@vrain.upv.es
}

\begin{document}

\maketitle

\begin{abstract}
  This paper presents the VRAIN-UPV MLLP's speech synthesis system for the SH1
  task of the Blizzard Challenge 2021. The SH1 task consisted in building a
  Spanish text-to-speech system trained on (but not limited to) the corpus
  released by the Blizzard Challenge 2021 organization. It included 5 hours of
  studio-quality recordings from a native Spanish female speaker. In our case,
  this dataset was solely used to build a two-stage neural text-to-speech
  pipeline composed of a non-autoregressive acoustic model with explicit
  duration modeling and a HiFi-GAN neural vocoder. Our team is identified as J
  in the evaluation results. Our system obtained very good results in the
  subjective evaluation tests. Only one system among other 11 participants
  achieved better naturalness than ours. Concretely, it achieved a naturalness
  MOS of $3.61$ compared to $4.21$ for real samples.
\end{abstract}

\section{Introduction}

The text-to-speech (TTS) field has witnessed a rapid progress over the recent
years. Particularly, deep learning based end-to-end
approaches~\cite{wang2017tacotron, Arik2017DeepVR, Shen18, Li2018CloseTH, Ren19,
yu2020durian, ren2021fastspeech} have gained popularity as they have shown to
bring improved naturalness and speech quality while significantly simplifying
the TTS training pipeline. This is not without saying that conventional speech
synthesis methods, like concatenative~\cite{541110, black1997automatically} or
statistical parametric speech synthesis (SPSS)~\cite{ZEN20091039} are still used
in many applications due to their advantages in robustness and efficiency.

With the text-to-speech field gaining momentum, the research community lacks of
proper automatic metrics and also of well-accepted test beds and methodologies
to evaluate speech naturalness, as opposed to other natural language processing
fields like ASR (Automatic Speech Recognition) or MT (Machine Translation).
Subjective listening tests adopting either the MUltiple Stimuli with Hidden
Reference and Anchor (MUSHRA) methodology or mean opinion score (MOS) ratings
have become the \emph{de facto} standard to assess and compare the speech
naturalness of TTS systems. A/B testing is also widely used when directly
comparing two alternative methods. To draw solid conclusions from the subjective
evaluations, these should include samples from all the considered models, and
the models should be trained under similar conditions. This implies a
significant effort, both in terms of research work and also computationally
speaking. Also, recruiting human participants for carrying out the subjective
evaluations is not always accessible (or affordable), especially for smaller
teams.

The Blizzard Challenge, organized annually since 2005, has the purpose of better
understanding and comparing different text-to-speech technologies applied to the
same provided training dataset. Since its inception, renowned IT companies and
institutions involved in text-to-speech research have been participating in the
different editions. In this year's challenge, the task SH1 consisted of building
a Spanish system from about 5 hours of studio-quality recordings from a native
female speaker. The organization allowed to further include up to a total of 100
hours of speech recordings from other sources for training the text-to-speech
models.

Our proposed system is composed of a non-autoregressive neural acoustic model with
explicit duration modeling and a GAN-based\footnote{GAN: Generative Adversarial
Network} neural vocoder. The acoustic model was initially developed from
ForwardTacotron\footnote{\url{https://github.com/as-ideas/ForwardTacotron}}, to
which we introduced several modifications based on different recently published
works. 
The acoustic model takes the phoneme sequence as inputs and generates an
intermediate speech representation (mel-spectrogram), which can be seen as a
lossy compressed version of the audio signal. Then, the vocoder model is
responsible of reconstructing the final speech waveform conditioned on the
mel-spectrograms. For the vocoder model, we used a public implementation of
HiFi-GAN~\cite{Su2020HiFiGANHD}, which is capable of producing high speech audio
quality significantly faster than real-time both on GPU and CPU. Both models are
described in detail in Sections~\ref{sec:model} and ~\ref{sec:vocoder},
respectively.

To summarize, the rest of the paper is organized as follows. First, the data
processing and the different tools used for this purpose are detailed in
Section~\ref{sec:data}. Then, the proposed forced-aligner autoencoder model used
to extract phoneme durations is presented in Section~\ref{sec:aligner}. The
acoustic TTS model architecture is described in detail in
Section~\ref{sec:model}. Section~\ref{sec:vocoder} introduces the GAN-based
vocoder model used in this work. The results of the subjective evaluation test
are given in detail and discussed in Section~\ref{sec:results}. Finally, some
conclusions are drawn in Section~\ref{sec:conclusions}.

\section{Data processing}
\label{sec:data}

This year, the organization provided participants with about 5 hours of
studio-quality recordings (after triming leading and trailing silence) from a
native Spanish female speaker and their corresponding transcriptions. The audio
samples were provided in 48kHz, PCM 16-bit format. Table~\ref{tab:data}
describes in detail the dataset released for the Blizzard Challenge 2021. As
mentioned above, the total duration of the recordings was computed after triming
leading and trailing silence from all samples.

\begin{table}[htpb]
\caption{Blizzard Challenge 2021 dataset}
\label{tab:data}
\begin{tabular}{llll}
Set & Samples & Number of words     & Duration (hours) \\\hline
SH1 & 4920    & 50.0 K              & 5.2              \\
SS1 & 10      & 0.1 K               & \textless 0.1
\end{tabular}
\end{table}

The text preprocessing procedure was carried out as follows. First, we
normalized the text transcriptions (lowercasing, removed special characters,
etc.) and extracted phoneme sequences from the normalized texts using the
well-known open-source speech synthesizer tool
eSpeak\footnote{\url{http://espeak.sourceforge.net}}, which includes a
pre-defined set of grapheme-to-phoneme (G2P) conversion rules for many languages
(including Spanish). The use of phoneme instead of grapheme sequences as inputs
to the acoustic model has probably just a minor effect (if any) in phonetically
simpler languages (e.g. Spanish, Italian or Portuguese, among others). On the
contrary, it can be certainly useful for phonetically more complex languages
(e.g. English), as it relieves the acoustic model from the task of inferring all
the different pronunciation rules from the training data.

Regarding the audio processing, we first resampled all the audio recordings to
22kHz and trimmed leading and trailing silence. Then, we extracted 100 bin log
magnitude Mel-scale spectrograms with Hann windowing, 50ms window length, 12.5ms
hop size and 1024 point Fourier transform. The spectrograms were finally min-max
normalized to lay within the $[0.0, 4.0]$ range.


Last, phoneme durations were extracted by training a separate forced-aligner
autoencoder model on the same dataset. This model is described next in
Section~\ref{sec:aligner}.

\section{Forced-aligner autoencoder model}
\label{sec:aligner}

Similarly to other non-attentive text-to-speech models with explicit duration
modeling~\cite{Ren19,yu2020durian,lancucki2020fastpitch,shen2020nonattentive,ren2021fastspeech},
our TTS acoustic model requires of pre-existing phoneme durations (in frames)
which are learnt during training. To extract phoneme durations, a monotonic
phoneme to frame alignment must be extracted from a separate model. Usually,
this is achieved by training an attention-based TTS model on the same data
(Tacotron, Transformer TTS~\cite{li2019neural}, etc.) or by using an external
forced alignment tool with pre-trained models like Montreal Forced
Aligner~\cite{mcauliffe2017montreal}. The former has been found to provide
slightly more convenient alignments as the aligner model is trained on the same
text-to-speech regression task as the acoustic model, as opposed to a pure
classification task.

Nevertheless, under suboptimal conditions (inaccurate transcriptions, noisy
recordings, small datasets, etc.) the training convergence of attention-based
TTS models is not guaranteed, particularly when the attention mechanism does not
constrain the alignments to be monotonic~\cite{He19,Battenberg20}. Also, the
training of the attention-based model is often computationally more expensive
than training the final TTS acoustic model. 

For these reasons, and inspired by Axel Springer Ideas Engineering's
DeepForcedAligner\footnote{\url{https://github.com/as-ideas/DeepForcedAligner}},
we developed a forced-aligner autoencoder model that makes use of an auxiliar
connectionist temporal classification (CTC) loss~\cite{graves2006connectionist}
to find the alignments between the spectrogram frames and the phoneme sequences.
This model is to be trained on the same speech data as the TTS model. Also, the
autoencoder framework is able to refine the CTC (or pure speech recognition)
alignments and make them more suitable for the TTS task. Last but not least, the
forced-aligner model provides enhanced robustness and significantly faster
convergence than attention-based TTS models.

\begin{figure}
  \centering
  \includegraphics[width=\linewidth,height=5cm]{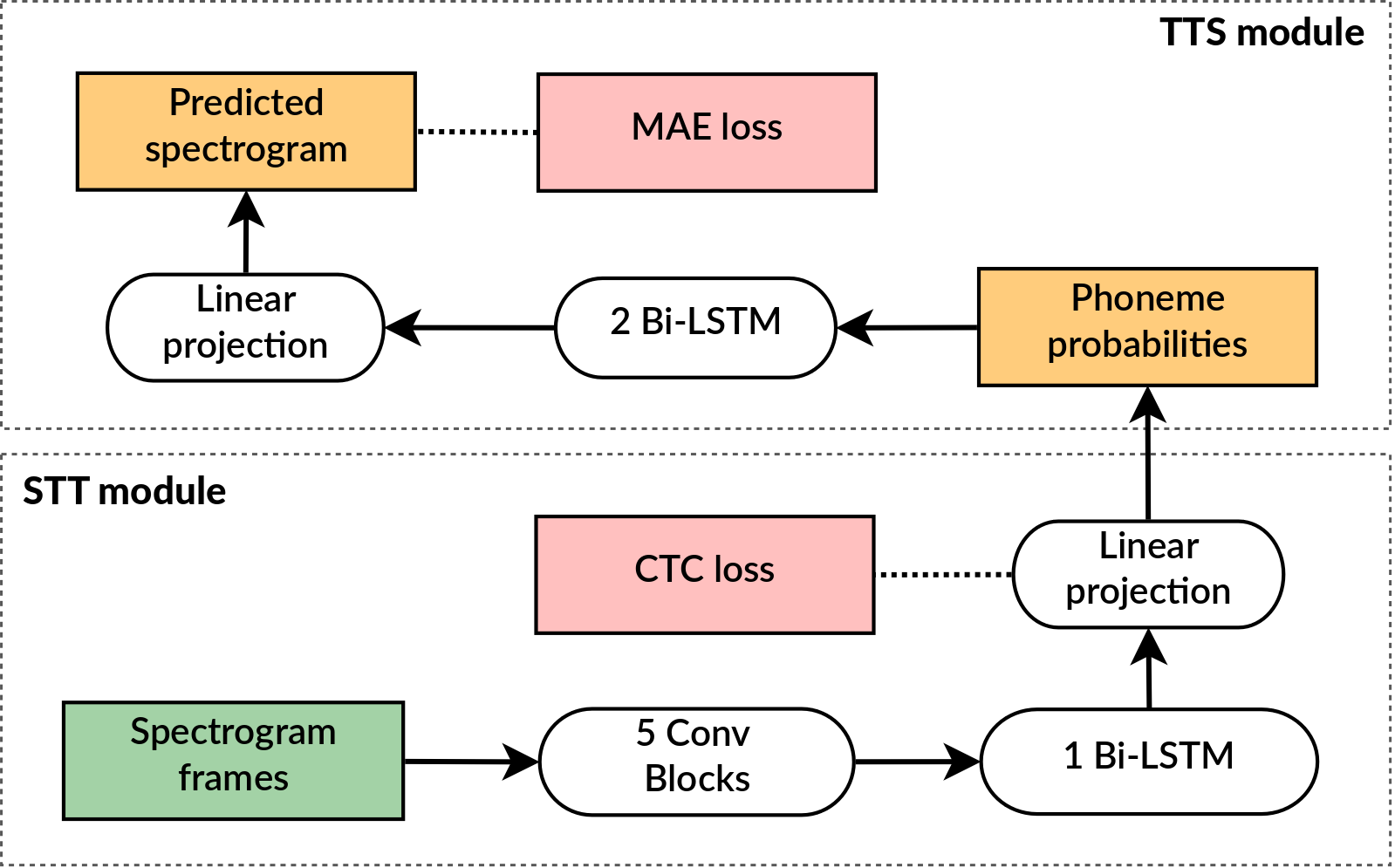}
  \caption{Forced-aligner autoencoder architecture overview. \label{fig:aligner}}
\end{figure}

Figure~\ref{fig:aligner} shows an overview of the proposed forced-aligner model.
It is composed of two interconnected modules following an autoencoder framework,
which are trained end-to-end with the help of an auxiliar CTC loss. The
speech-to-text (STT) encoder module is a simple speech recognition model. The
input spectrogram frames are passed through a stack of 5 1-D convolutional
layers, followed by batch normalization and ReLU activations. The output of the
last convolution layer is then processed by a single-layer bi-directional LSTM,
followed by a final linear projection layer with softmax activations. An
auxiliar CTC loss computed over the ground-truth phoneme sequences is used on
the softmax outputs to help the convergence of the STT module. Then, a simple
text-to-speech (TTS) module plays the decoder role of the autoencoder framework.
This module takes the STT softmax outputs (phoneme class probabilities) as
inputs, and aims to reconstruct the original spectrogram frames. It is composed
of 2 bi-directional LSTM layers followed by a linear projection to the
spectrogram dimension. The mean absolute error (MAE) between the ground-truth
and the generated spectrograms is backpropagated through the entire autoencoder
model helping refining the STT alignments for the text-to-speech task.

To extract phoneme durations, we discard the decoder (TTS) module and use the
speech-to-text encoder to calculate phoneme posteriors. Then, we use Dijkstra's
algorithm to find the most likely monotonic path through the sequence of phoneme
probabilities.

\section{Acoustic model}
\label{sec:model}

Recently, autoregressive neural TTS models have shown state-of-the-art
performance in terms of enhanced speech naturalness~\cite{Shen18, li2019neural}.
Despite the high synthesis quality of these models, they generally show a lack
of robustness at inference time, particularly when the attention mechanism does
not constrain the alignments to be monotonic. Attention failures cause
mispronounciation, skipping or repeating words, and even totally unintelligible
speech when the attention mechanism collapses.

To overcome such limitations, non-autoregressive parallel TTS models with
explicit duration modeling have been proposed~\cite{shen2020nonattentive,
lancucki2020fastpitch} with similar success. These models do not only
solve the undesired attention failures, but are also significantly faster
than their autoregressive counterparts.

\begin{figure}
  \centering
  \includegraphics[width=\linewidth]{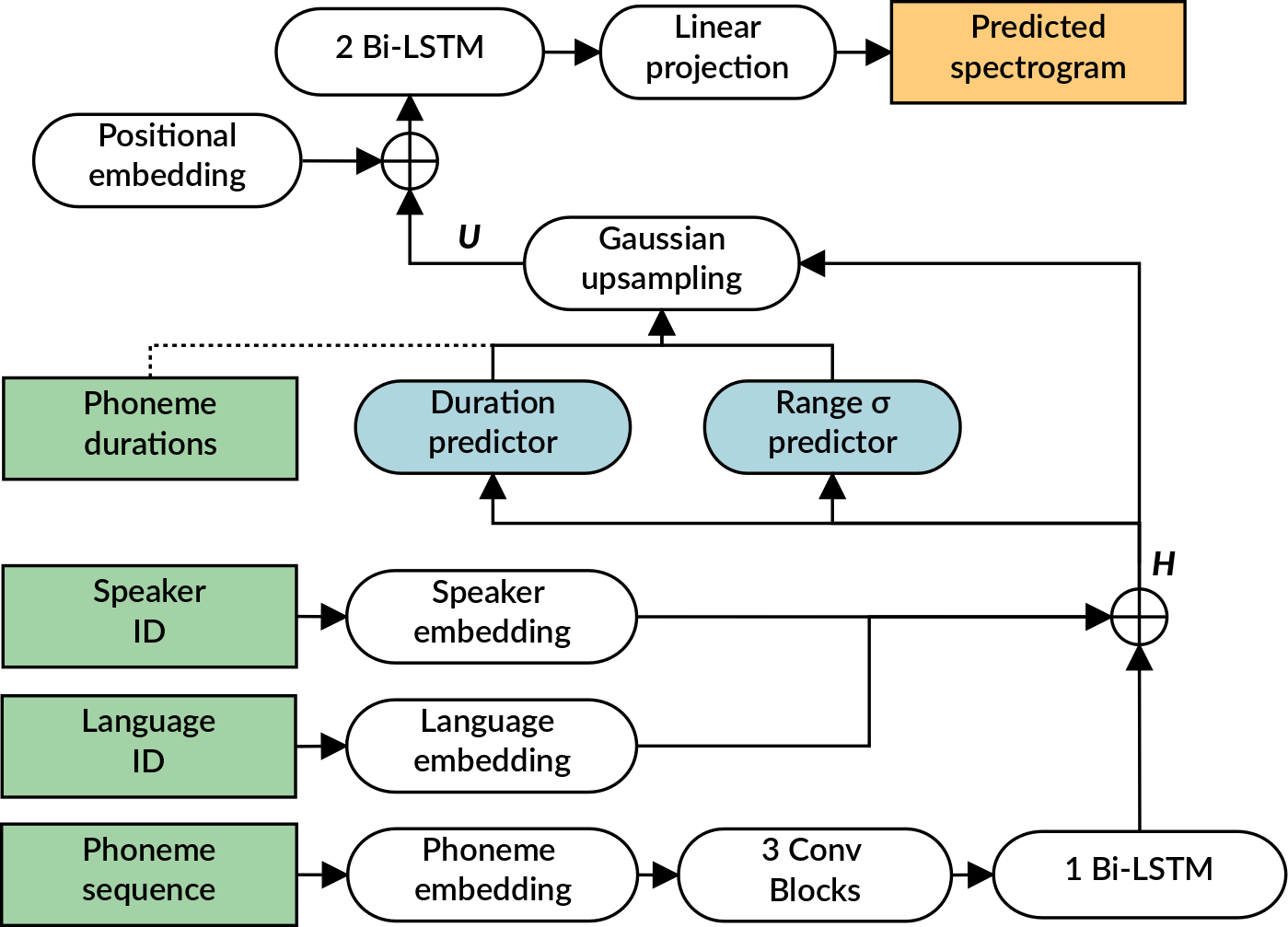}
  \caption{Proposed TTS acoustic model architecture. Dashed lines are training-only
    connections. \label{fig:model}}
\end{figure}

The proposed model was initially based on ForwardTacotron%
\footnote{\url{https://github.com/as-ideas/ForwardTacotron}}, an open-source
non-autoregressive variant of Tacotron~\cite{wang2017tacotron} inspired on
parallel TTS models like FastSpeech~\cite{Ren19} or DurIAN~\cite{yu2020durian}.
The original architecture was composed of two PreNet bottleneck layers, a CBHG
encoder~\cite{wang2017tacotron}, a variance duration predictor similar to
~\cite{Ren19}, a 2-layer Bi-LSTM decoder and a convolutional residual PostNet as
in~\cite{Shen18}. However, following more recent works, we have introduced
several modifications to this original architecture. The final modified
architecture is depicted in Figure~\ref{fig:model}, and the introduced
modifications are described in detail next.

First, we replaced the encoder module with the simplified architecture proposed
in Tacotron-2~\cite{Shen18} shown at the bottom of Figure~\ref{fig:model}. The
encoder module consists of learned 512-dimensional phoneme embeddings that are
passed through a stack of three 1-D convolutional layers, followed by batch
normalization and ReLU activations. The output of the last convolutional layer
is processed by a single bidirectional LSTM layer to generate the encoder hidden
states. These states are later expanded attending to phoneme durations and
consumed by a decoder to generate the spectrogram frames.

Second, we replaced the vanilla upsampling through repetition (also known as
length regulator module) in favor of the Gaussian upsampling approach recently
proposed in~\cite{shen2020nonattentive}, which has been shown to improve speech
naturalness. After the upsampling, we also add Transformer-style sinusoidal
positional embeddings of a frame position with respect to the current phoneme as
in~\cite{elias2020parallel}. We also discarded the recurrent layer of the variance
predictor module as we empirically found improved robustness on phoneme duration
predictions, in line with~\cite{Ren19}.

Finally, we discarded the convolutional residual PostNet as we found it not
to bring any noticeable improvements on the spectrogram reconstruction task when
using a bi-directional LSTM decoder.

The text-to-spectrogram models are trained using a combination of the
$\mathcal{L}1$ loss and the \textit{structural similarity index measure} (SSIM)
between the predicted and the target spectrograms, and \textit{Hubber loss} for
logarithmic duration prediction~\cite{Vainer2020}.

\section{Vocoder model}
\label{sec:vocoder}

The public HiFi-GAN implementation%
\footnote{\url{https://github.com/jik876/hifi-gan}} is used for reconstructing
the audio waveform conditioned on the generated spectrograms. We choose the
HiFi-GAN V1 ($h_u = 512$), which is the larger HiFi-GAN model proposed in the
original paper and brings the better audio quality compared to the reduced V2
and V3 models~\cite{Su2020HiFiGANHD}. It was trained only on the audio
recordings provided by the organization. The training procedure comprises two
steps. Initially, the vocoder model is trained on the extracted ground-truth
spectrograms for 500K steps. Then, after the acoustic model is trained,
ground-truth aligned (GTA) spectrograms are generated for the training dataset
and the HiFi-GAN model is fine-tuned on the acoustic model outputs for an
additional 100K steps, which helps reducing the artifacts induced by the
missmatch between the vocoder training and inference conditions and brings
slightly better audio quality for the TTS task.

\section{Subjective results}
\label{sec:results}

A total of 12 participating teams submitted their generated test samples to be
evaluated in the subjective listening test. Table~\ref{tab:resultssec} describes
the different aspects considered for the subjective evaluation test. Our system
was assigned the letter J, while R corresponds to the original audio recordings.

\begin{table}[]
  \caption{Subjective evaluation sections}
  \begin{tabular}{lll}
  Section          & Aspect               & Metric \\\hline
  Sections 1 and 2 & Speaker similarity   & MOS (1-5) \\
  Sections 3 and 4 & Naturalness          & MOS (1-5) \\
  Section 5 (Sharvard) & Intelligibility  & WER \% \\
  Section 6 (SUS)  & Intelligibility      & WER \%
  \end{tabular}
  \label{tab:resultssec}
\end{table}

The listeners participating in the subjective test could be divided into three
different groups: 

\begin{itemize}
\item SP: Paid participants (native speakers of Spanish)
\item SE: Volunteer speech experts (self-identified as such)
\item SR: Rest of volunteers
\end{itemize}

\subsection{Naturalness}
\label{sec:naturalness}

In the speech naturalness test (sections 3 and 4), listeners listened to one
sample and chose a score which represented how natural or unnatural the sentence
sounded on a scale of 1 (completely unnatural) to 5 (completely natural).
Building systems that can achieve a high degree of naturalness is probably the
most challenging task when it comes to generate synthetic speech, and thus the
speech naturalness MOS has become the main metric to evaluate speech synthesis
quality over the last few years.

\begin{figure}
  \caption{\label{fig:naturalness} Naturalness MOS for all participants
  (all listeners)}
  \includegraphics[width=\linewidth]{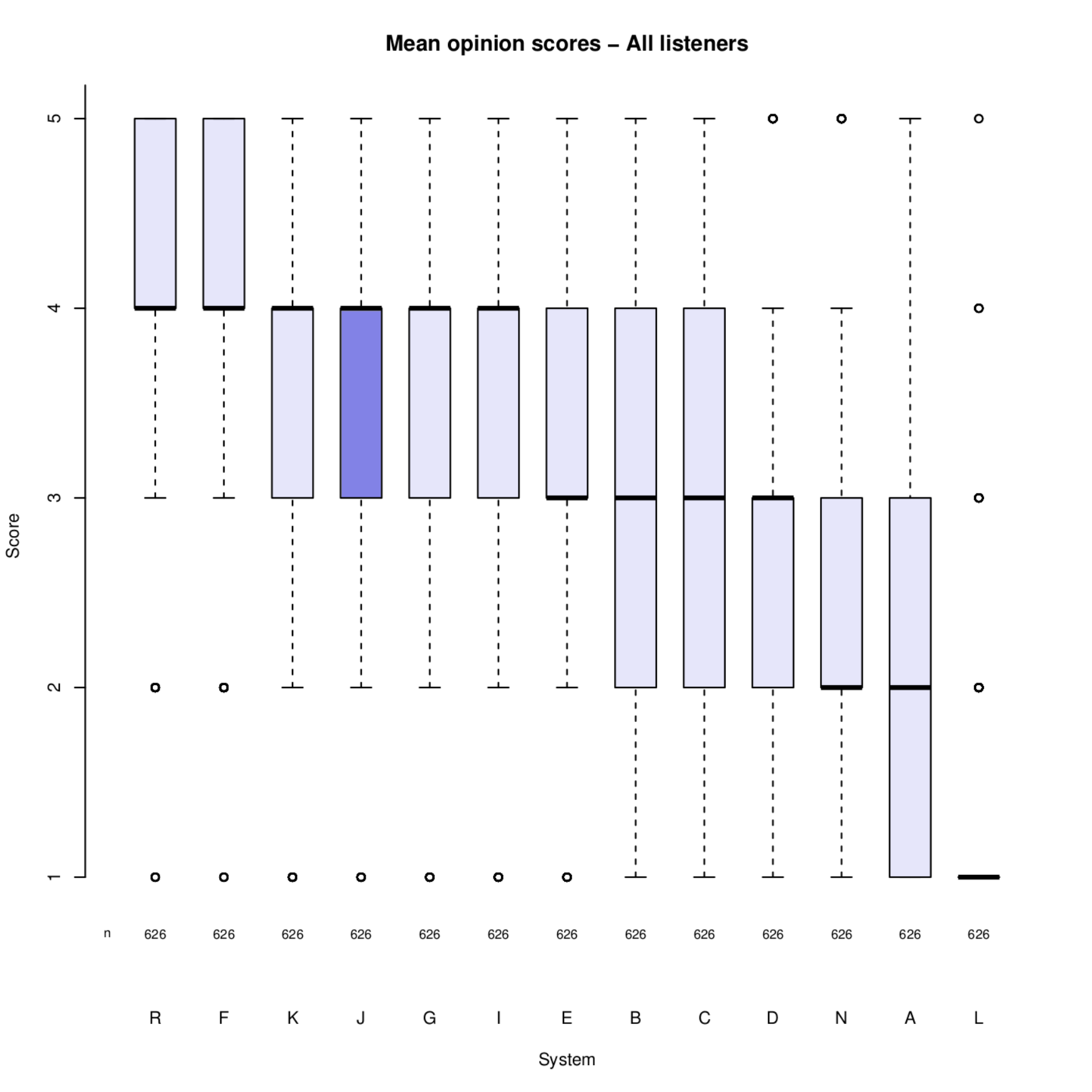}
\end{figure}

The boxplot evaluation results of all systems on speech naturalness from all
listeners is showed in Figure~\ref{fig:naturalness}. In this case, system F
performed clearly better than other systems. Our system was scored with a
naturalness MOS of $3.61$ in terms of speech naturalness, while the real
recordings were scored with $4.21$. As can bee seen in
Figure~\ref{fig:naturalness}, our system performance is comparable to that of
systems K, G and I, all of them ranked in a second position just after system F.
We believe this result is very positive, especially when considering our system
was trained with a limited amount of data ($\sim$5 hours) to what is common in
two-stage neural TTS pipelines (20 hours or more)~\cite{ljspeech17,
Veaux2017CSTRVC}.

\subsection{Speaker similarity}
\label{sec:sim}

In the speaker similarity test (sections 1 and 2), listeners could play 2
reference samples of the original speaker and one synthetic sample. They chose a
response that represented how similar the synthetic voice sounded to the voice
in the reference samples on a scale from 1 (sounds like a totally different
person) to 5 (sounds like exactly the same person).

\begin{figure}
  \caption{\label{fig:sim} Speaker similarity scores for all participants (all
  listeners)}
  \includegraphics[width=\linewidth]{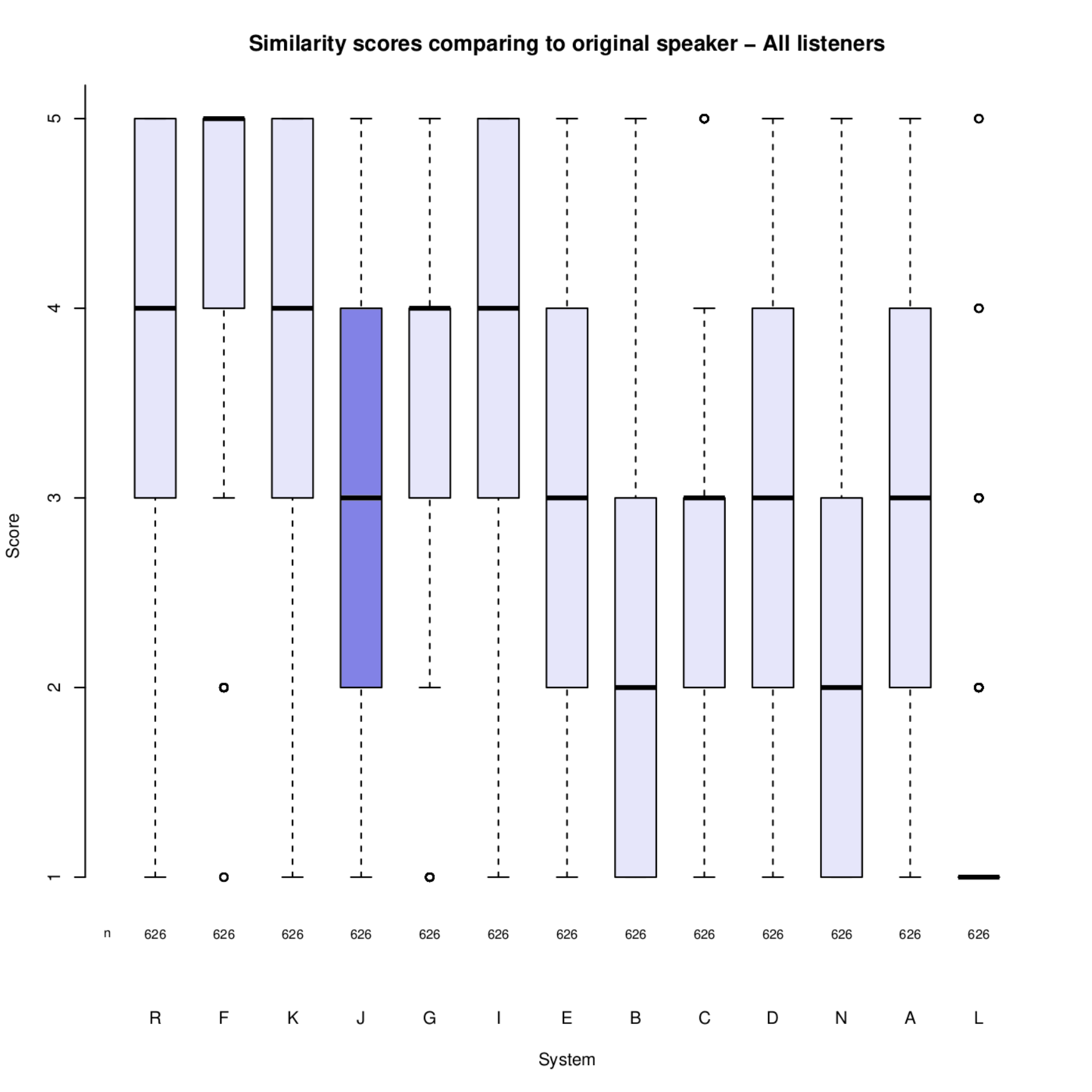}
\end{figure}

The boxplot evaluation results of all systems on speaker similarity from all
listeners is showed in Figure~\ref{fig:sim}. System F, K and I performed better
than other systems. Despite the fact that our system (J) was only trained with
the data provided by the Blizzard Challenge 2021 organizers, it was scored with
a speaker similarity MOS of $3.29$ compared to $4.07$ from the original
recordings (R).

\subsection{Intelligibility test}
\label{sec:int}

Finally, an intelligibility test was carried out in sections 5 and 6. The goal
of this test was just to determine whether or not the synthetic speech was
understandable. Listeners heard one utterance in each part and typed in what
they heard. Listeners were allowed to listen to each sentence only once. The
sentences were specially designed to test the intelligibility of the synthetic
speech: the sentences and the reference natural recordings for section 5 came
from the Sharvard corpus, while the SUS for section 6 were kindly provided by
TALP-UPC and Aholab-EHU research laboratories.

\begin{figure}
  \caption{\label{fig:intsharvard} Intelligibility Word Error Rates (WER) for the Sharvard intelligibility test}
  \includegraphics[width=\linewidth]{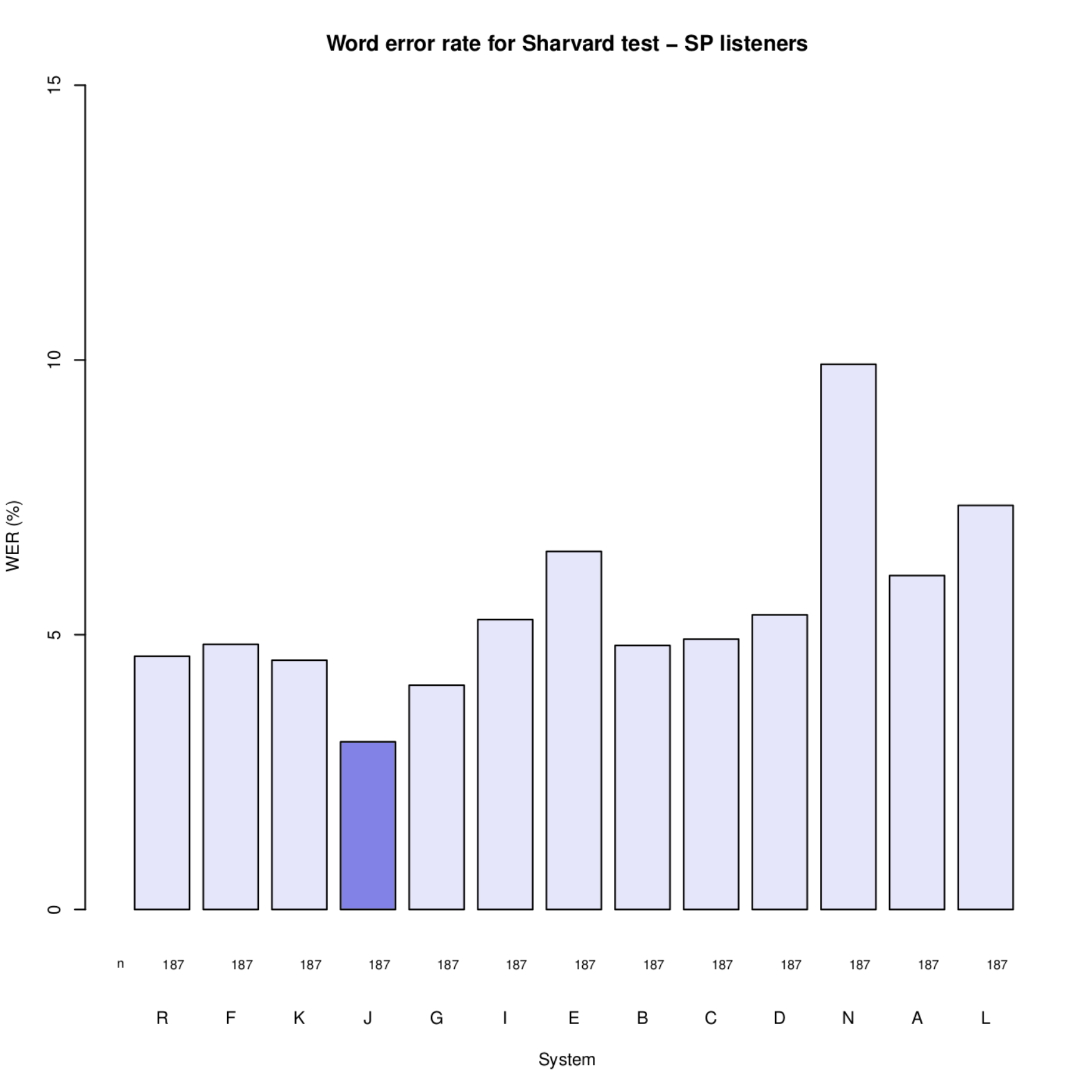}
\end{figure}

\begin{figure}
  \caption{\label{fig:intsus} Intelligibility Word Error Rates (WER) for the SUS intelligibility test}
  \includegraphics[width=\linewidth]{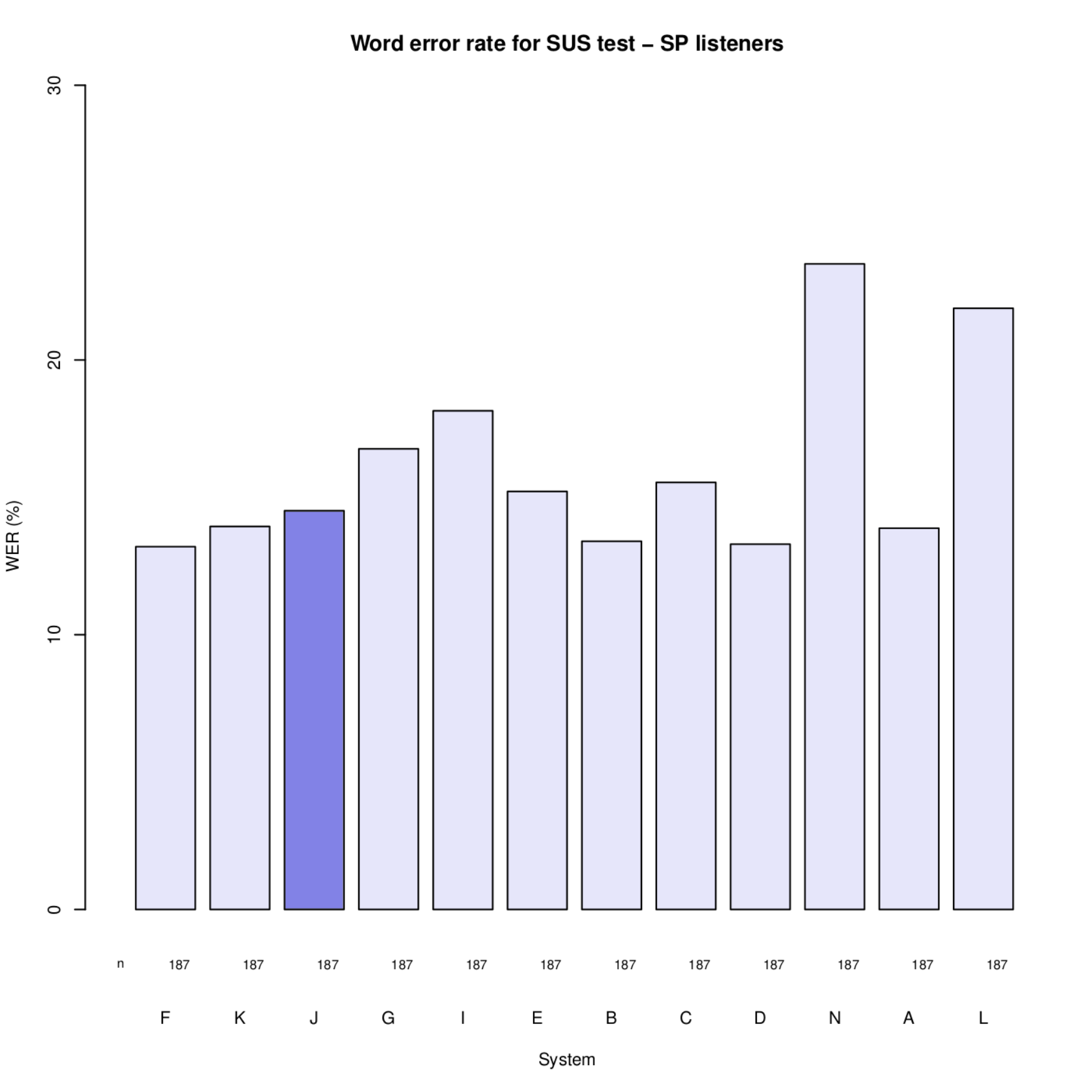}
\end{figure}

Figure~\ref{fig:intsharvard} shows the Word Error Rates (WER) of all teams for
the Sharvard intelligibility test, where our system achieved the lowest
transcription error ($3.0$\%). This result emphasizes the good performance of
our model regarding word pronunciation even though it was trained with a limited
amount of speech data. Figure~\ref{fig:intsus} shows the WER for the SUS
intelligibility test, where our system performs similarly to the best performing
systems.

\section{Conclusions}
\label{sec:conclusions}

In this work we have presented our proposed two-stage neural text-to-speech
system for the Blizzard Challenge 2021. Both the acoustic and the vocoder models
were trained using only the data provided by the organization (close to 5 hours
of studio-quality recordings from a Spanish female native speaker). In the
subjective listening tests, our system (identified as J) performance in terms of
speech naturalness MOS was ranked in second position along with 3 other systems.

\section{Acknowledgements}

The research leading to these results has received funding from the Government
of Spain's research project Multisub (ref. RTI2018-094879-B-I00,
MCIU/AEI/FEDER,EU).

\bibliographystyle{IEEEtran}

\bibliography{blizzard21}

\end{document}